\documentclass[11pt]{article}
\usepackage{amssymb,amsmath}
\usepackage[dvips]{graphics}
\usepackage[dvips]{rotating}
\usepackage{epsfig}

\begin{document}
\title{Optimizing the Linear Collider Detector for the Measurement of the Higgs to Charm Branching Ratio\footnote{Talk presented by A.C. at the Linear Collider Workshop at the University of Chicago, January 7-9, 2002.}}

\author{ TOSHINORI ABE and AARON S. CHOU \\
Stanford Linear Accelerator Center, 2575 Sand Hill Road,\\
Menlo Park, CA, USA\footnote{Work supported in part by Department of Energy contract DE-AC03-76SF00515.}}
\date{}
\maketitle

\begin{abstract}
Several different vertex detector designs for the Linear Collider 
Detector (LCD)
are evaluated in the context of measurements of the branching ratio
of the standard model Higgs particle decaying into charm/anti-charm final
states.  Fast Monte Carlo simulations are used to model the detector
and neural network-optimized flavor tagging is used to perform
the measurements.  These tools are used to study the effects of pixel
resolution, material thickness, and inner layer radius on the 
flavor tagging efficiency-purity curve and ultimately on the branching
ratio measurement error.
\end{abstract}

\section{Introduction}
A worldwide effort is now underway to develop a future Linear Collider 
and a corresponding Linear Collider Detector (LCD).  Compromises may
have to be made in the design of such a detector between its desired
performance and other considerations such as the restrictions imposed
by beam-related backgrounds in the interaction region as well as the
total cost of the detector.  In this paper, we use Monte
Carlo simulations to model the performance of various detector
configurations in the contexts
of flavor tagging and sensitivity in the measurement of the branching
ratio of a 120 GeV Higgs boson to charm/anti-charm final states.  Such
a measurement would help determine whether the condensation of a single
Higgs boson is sufficient to generate the masses of particles.  The
dependence of the measurement precision on the various detector parameters
used provides useful data for optimizing the final design of the detector.

The branching ratios $BR(H^0\rightarrow c\bar c)$ and 
$BR(H^0\rightarrow b\bar b)$ may be simultaneously measured using the SLD
technique \cite{Abe:1998ez} for measuring the the analogous quantities 
$R_b$ and $R_c$ in $Z^0$ decays.  The branching ratios may be measured
by performing flavor tagging and then counting the number of
$b$-tagged, $c$-tagged, and untagged decays.  These data can then be
unfolded using the flavor tagging efficiency matrix.  

Efficient and pure flavor
tagging requires precise impact parameter resolution in order to distinguish
heavy flavor decay tracks from fragmentation tracks from the primary
interaction point (IP).  This tracking resolution may be written as
$\sigma_0 \oplus \sigma_{MS}/(p \cdot \sin^{3/2}\theta)$  for a cylindrically
symmetric detector where $\theta$ is measured from the $z$ axis and $p$ is
the particle's momentum.  $\sigma_0$
is the asymptotic resolution at infinite momentum and it scales
linearly with the detector's single hit resolution.  For finite momentum,
the total resolution is usually dominated by the second term where
$\sigma_{MS}$ includes the effects of multiple Coulomb scattering
and thus scales as the square root of the detector layer thickness. 
Both errors scale
roughly linearly with the inner radius of the detector which is effectively
the extrapolation distance from the innermost hit inward to the IP.
The effects of variations of these three detector parameters on flavor
tagging are described below.  Most of the same assumptions are made
as in a previous study of the same topic by the University of Oregon 
group\cite{Brau:2000dq}, and our results are very consistent with theirs.
However, the current study explores several additional detector configurations.
 
\section{The Analysis Tools}
The various vertex detector configurations are simulated using the
LCDTRK program\cite{Schumm:1999es} which, given an input detector geometry,
generates smearing tables for the track parameter error matrices.  
Non-Gaussian measurement errors are not modelled.  These smearing tables
are input into the LCDROOT\cite{Abe:2001up} package for the event
generation and fast MC detector simulation.  

To calibrate the fast MC, the SLD tracking resolution is simulated by
replacing the tracking chambers of the small 
detector (SDMAR01)\cite{Abe:2001nr} with
a model of the SLD tracking chambers \cite{Abe:vxd3} \cite{Hildreth:1994yq}
\cite{Fero:1995pv} .  This model is tuned to reproduce the measured SLD
track position resolution of 
$\sim (9 \oplus 33/((p/{{\rm GeV}) \cdot \sin{\theta}^{3/2})} \ \mu$m.   
For this tuning, the beampipe thickness is increased from $0.52\% \ X^0$ to 
$0.65\% \ X^0$ in order to accomodate the increased measurement error from
Rutherford scattering tails.  The magnetic field for the drift chamber is
set to 0.6 T and interaction point (IP) measurement errors
$(4\cdot 4\cdot 20) \mu$m are modelled by smearing the MC IP position. 
20K $Z^0$ pole hadronic decay events are
generated and the LCDJetFinder and LCDVToplGhost packages in LCDROOT are
used to find jets using the Durham algorithm, and to perform topological
vertexing \cite{Jackson:zvtop} \cite{Abe:bsghost} within the jets.  

 \begin{figure}[htb]
 \centering
 \hspace{1cm}
 \begin{minipage}{.40\textwidth}
 \rotatebox{-90}{\resizebox{\textwidth}{!}{\includegraphics{ctagsldlcd.epsi}}}
 \end{minipage}
 \hfill
 \begin{minipage}{0.4\textwidth}
 \rotatebox{-90}{\resizebox{\textwidth}{!}{\includegraphics{btagsldlcd.epsi}}}
 \end{minipage}
 \caption{\label{F:fmcsld}The efficiency-purity curves of the LCDROOT fast MC jet $b$-tag (left) and $c$-tag (right) for the SLD detector.  The `ghost track' algorithm is used for vertexing.}
\bigskip
\bigskip
\begin{minipage}{\textwidth}
\resizebox{\textwidth}{!}{\includegraphics{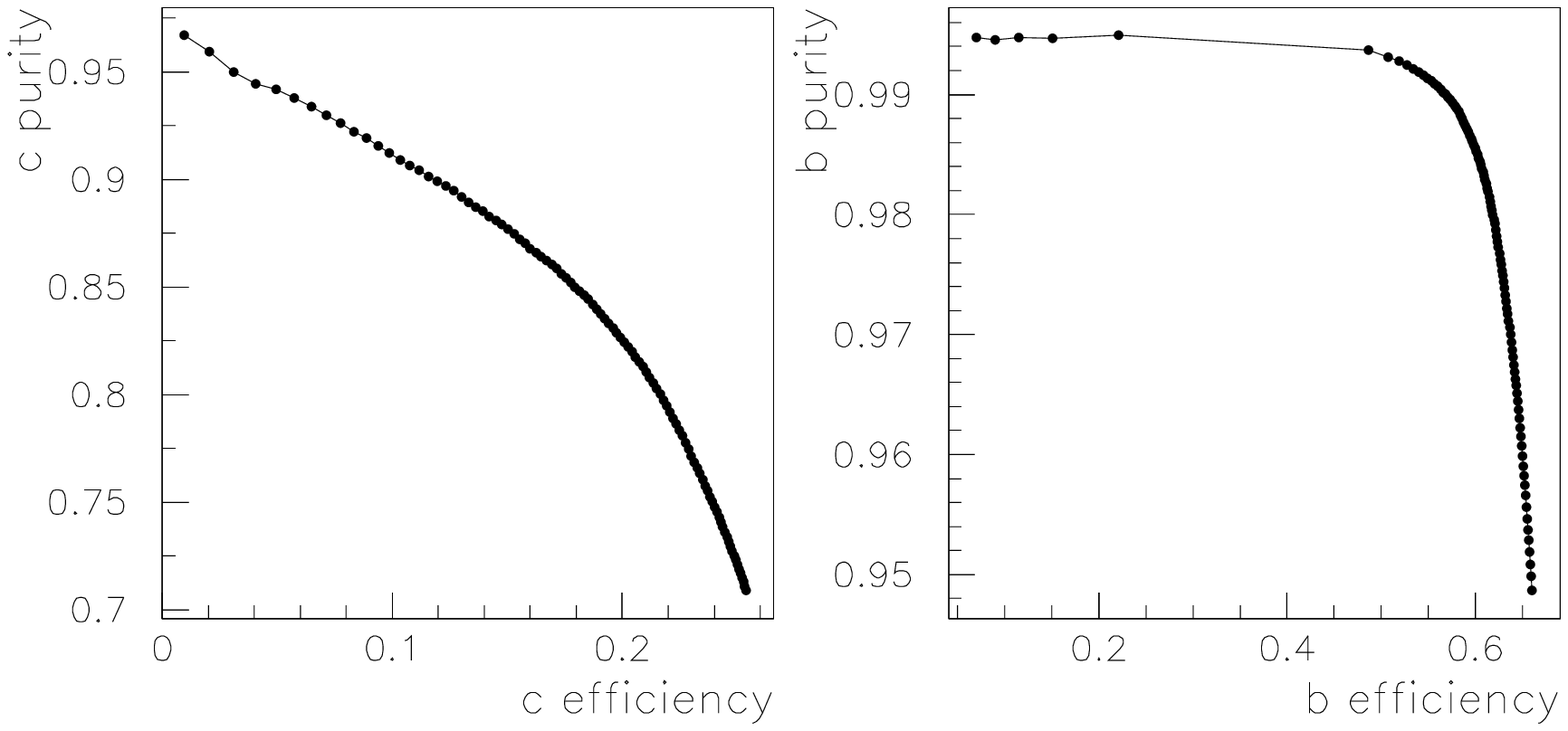}}
 \caption{\label{F:effpur}The efficiency-purity curves for the full MC jet $c$-tag (left) and $b$-tag (right) for the SLD detector.  The `resolubility' algorithm is used for vertexing.}
 \end{minipage}
 \end{figure}

A jet flavor tag analogous to the SLD method\cite{Rowson:2001cd} is
then developed.  The tag proceeds in three stages,
each of which is optimized with the Stuttgart neural network trainer.
First, the best candidate $b$ vertex is selected
from the vertices found in the jet.  This selection is based on
the vertex flight distance, the flight distance normalized by the vertex
measurement error, the angle of the vertex flight direction with respect
to the jet axis, the number of tracks in the vertex, and the raw vertex
mass computed from the track momenta.  

Next, tracks are attached to the $b$ vertex using the following criteria:
the transverse distance of the track to the vertex flight axis at
the point of closest approach $(POCA)$, the longitudinal distance 
of the $POCA$ on the vertex flight axis to the IP, this longitudinal
distance divided
by the vertex flight distance, the angle of the track momentum with 
respect to the vertex flight direction, and the track impact parameter
normalized by its measurement error.  The vertex kinematics are then
recalculated using all of the attached tracks.  
 
Finally, the jet flavor is identified using the $P_T$-corrected vertex mass,
the vertex momentum, the vertex flight distance, the flight distance
normalized by the vertex measurement error, the number of vertices found,
and the number of 1-prong vertices found.  The results of the jet flavor 
tag are shown in figure~\ref{F:fmcsld}.  These efficiency-purity curves
are very similar to the the flavor tag results of the full SLD MC shown
in figure~\ref{F:effpur}.  The fast MC $b$ tag has slightly lower purity
perhaps because the `ghost track' vertexing algorithm used in the fast MC
tag instead of the `resolubility' vertexing algorithm used in the SLD tag.
The `ghost' algorithm is more susceptible to creating vertices from
poorly measured fragmentation tracks, thus incorrectly increasing the
vertex mass.  The discrepancy is very small however, and we conclude
that the fast MC using LCDTRK and LCDROOT is sufficiently good for our
purposes.

\section{Flavor Tagging in Higgstrahlung Events}
To measure the various Higgs branching fractions, we look at 
Higgstrahlung events in which the final state $Z^0$ is identified
by reconstructing its decay into $e^+ e^-$ or $\mu^+ \mu^-$ 
and a 120 GeV Higgs is identified through its recoil mass.  The efficiency
of the recoil mass technique to tag a 120 GeV Higgs is estimated
to be $\sim 31\%$ with $\sim 44\%$ impurity from $Z^0 Z^0$ events  
due to degradation of the beam energy constraint from beamstrahlung and 
initial state radiation.  To reduce the $Z^0\rightarrow c\bar c$ background,
a cut of $|\cos{\theta}| < 0.6$ is imposed on the jet axes to remove
a large fraction of the polarized $Z^0$'s produced by the $80\%$ polarized
electron beam.  The resulting Higgs efficiency is $\sim 19\%$ with
$75\%$ purity.  Since only leptonic decays of the $Z^0$
are used, the Higgs decay jets can be selected unambiguously in this
study, but only $\sim 6 \%$ of the Higgstrahlung events can then be used. 
However, jets from Higgs and hadronic $Z^0$ decays can be separated some
fraction of the time, and so the fraction of usable events in a full
analysis can very likely be increased by a factor of 4.  Assuming
a 60 fb Higgstrahlung cross-section, the total number of usable tagged
Higgs decays in a 500 fb$^{-1}$ sample is then estimated to be:
\begin{equation} 
N_H \approx 500 \ {\rm fb}^{-1} \cdot 60 \ {\rm fb} \cdot 0.19 \cdot 0.06 \cdot 4 \approx 1370.
\end{equation}

For each detector configuration, 20K Higgstrahlung events are generated
in which the $Z$ is forced to decay into $e^+ e^-$ or $\mu^+ \mu^-$ and
the 120 GeV Higgs decays hadronically.  Since the branching ratio 
$BR(H\rightarrow c\bar c)$ is expected to be low, to get a good
statistical sample of charm jets, an additional 20K Higgstrahlung events
are generated in which the Higgs are now forced to decay to $c \bar c$.  
The magnetic field is set to 5 Tesla.

The IP position measurement is modelled to have $(4\times 4\times 6) \mu$m
resolution.  These numbers are motived by the SLC IP position
measurement\cite{Chou:2001gj} which is composed of measurements
of the beam position averaged over many events, and an event-by-event
measurement of the longitudinal position along the beam of the primary 
vertex.  The SLC beam position is measured by putting fragmentation
tracks from 30 consecutive hadronic events into a common 2D vertex in
order to average over the jet axis distribution in $\phi$.  At luminosities of
$10^{34}$ cm$^{-2}$ sec$^{-1}$
the event rate for $W^+W^-$ and $Z^0\gamma$ events at the Linear Collider
should be comparable to the $Z^0$ event rate ($\sim 200$/hour) at the SLC,
and this technique should still work provided that the beam does not move by
more than a few microns per hour.  The SLC longitudinal IP position is 
measured event-by-event using the point of closest approach to the beam
of the median track in the track-to-beam impact parameter distribution of
tracks in the event.  The
resolution of this technique is expected to be of the same order as the
average track position resolution.  The changes in the longitudinal
IP resolution when using
different detector configurations have been neglected and so the changes
in the branching ratio measurement precision for the most extreme detector
configurations in table~\ref{T:trackres} may be slightly underestimated.

 \begin{figure}[htb]
 \centering
 \begin{minipage}{.44\textwidth}
 \resizebox{\textwidth}{!}{\includegraphics{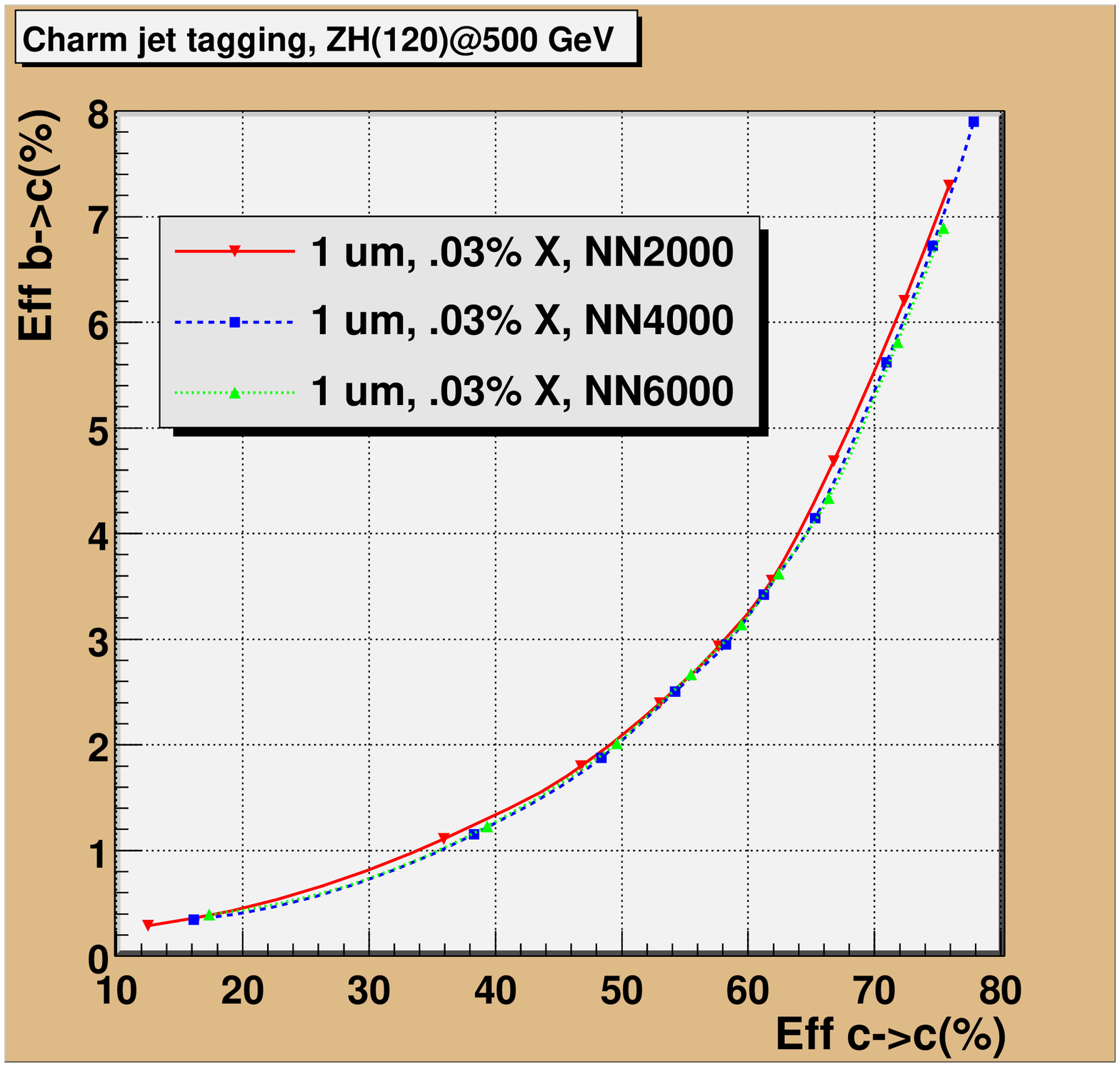}}
 \caption{\label{F:nncycle} The $b\rightarrow c$ mistag rate versus the $c\rightarrow c$ correct tag rate of the jet flavor tag for different numbers of neural network training cycles.}
 \end{minipage}
 \hfill
 \begin{minipage}{.44\textwidth}
 \resizebox{\textwidth}{!}{\includegraphics{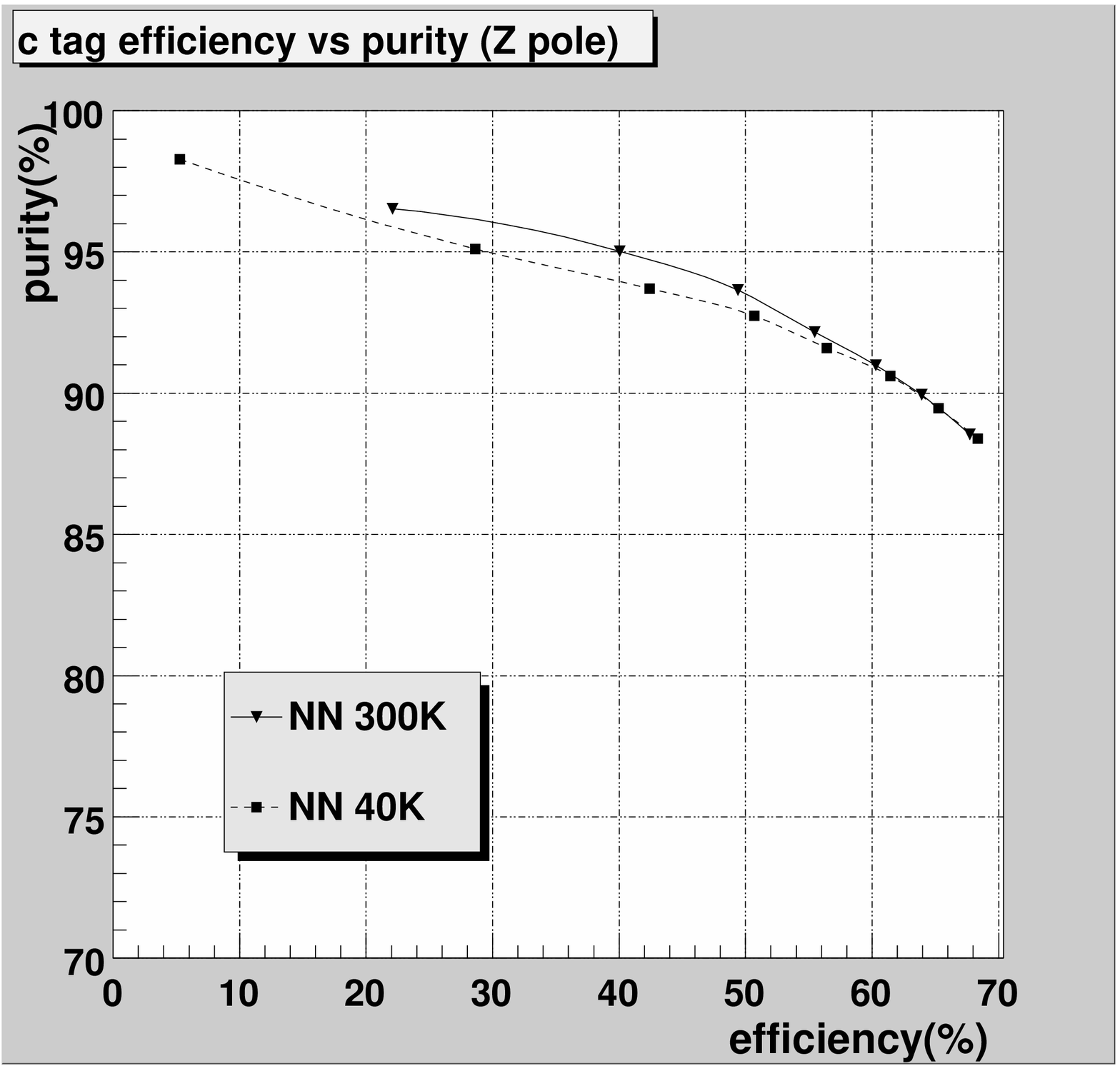}}
 \caption{\label{F:nnevents} The efficiency-purity curve of the jet flavor tag for neural networks trained with 300K $Z^0$ events and with 40K $Z^0$ pole events.} 
 \end{minipage} 
 \end{figure}

The vertex selection and track attachment neural networks are trained
only once for the standard SDMAR01 configuration\cite{Abe:2001nr},
but the jet flavor
tagging neural net is retrained using these 40K events for each 
new detector configuration.  20K of the events are used for training and 
20K for validation to avoid overtraining.  The same training regimen
is used for each configuration:  1000 cycles of standard backpropagation
with the learning rate set to 0.2 and then 1000 cycles each of
backpropagation with momentum 0.5 and the learning rate set to 0.1, 0.05,
and 0.01.  Studies showed that additional training beyond this basic
regimen yield only marginal improvements in the flavor tagging.  Small
improvements could also be made using a 300K event sample for the
training, but working with such large sample sizes was not feasible for
this quick study.

Since each hadronic Higgs decay produces both a quark jet and an anti-quark 
jet, there are two chances to tag the decay type.  The event level flavor
tag may be defined as follows.  If either or both jets are $b$-tagged, then
the event is $b$-tagged.  If either or both jets are $c$-tagged and neither
is $b$-tagged, then the event is $c$-tagged.  Otherwise, the event is
untagged and considered to be a decay into light quarks.  For example,
the event flavor tagging efficiencies $\mathbf{E}\equiv E_{ij}$ for
identifying flavor $i$ as flavor $j$ may be expressed in terms of the jet
flavor tagging efficiencies $\epsilon_{ij}$ as:
\begin{eqnarray}
 E_{cc} &=& \epsilon_{cc}^2 + 2 \epsilon_{cc} \cdot \epsilon_{cu} \nonumber \\
\label{E:Ecc}  &=& 2 \cdot \epsilon_{cc}\cdot(1-\epsilon_{cb})-\epsilon_{cc}^2 \\
 E_{bc} &=& \epsilon_{bc}^2 + 2 \epsilon_{bc} \cdot \epsilon_{bu} \nonumber \\
\label{E:Ebc}  &=& 2 \cdot \epsilon_{bb}\cdot(1-\epsilon_{bb})-\epsilon_{bc}^2 
\end{eqnarray}
Here, the unitarity of the jet flavor tagging efficiency matrix is used 
for the second equality and $i$ and $j$ take on values of $b$, $c$ and
$u$ for untagged.  Values of $\epsilon_{bb}\approx 0.6$ and
$\epsilon_{cb}\approx 0.02$ are typical.  In these formulae, the effects
on flavor tagging due to
energy correlations between the two jets in a decay are neglected.
However, since the $BR(H\rightarrow c\bar c)$
measurement is dominated by the $H\rightarrow b\bar b$ background, these
correlations can only improve the measurement by reducing $E_{bc}$ by 
making it less likely that the two $b$ jets are both tagged as $c$.  

\section{Measuring the Branching Fractions}
The branching ratios $\mathbf{BR_H} \equiv (BR_{H\rightarrow b\bar b}, BR_{H\rightarrow c\bar c}, BR_{H\rightarrow u\bar u})$ may be extracted by counting the
number of event level flavor tags of each flavor $\mathbf{T} \equiv (T_b, T_c, T_u)$
by solving the matrix equation:
\begin{equation}
\mathbf{T} = \mathbf{E} \cdot (\mathbf{BR_H} \cdot N_H + \mathbf{BR_Z} \cdot N_Z)
\end{equation}
to get:
\begin{equation}
\Delta \mathbf{BR_H} = \mathbf{E}^{-1}\cdot (\mathbf{T}-\mathbf{E} \cdot \mathbf{BR_Z} \cdot N_Z) / N_H 
\end{equation}
Here, $N_H$ and $N_Z$ are
the numbers of Higgs and $Z^0$'s in the Higgs-tagged sample,
and the $Z^0$ decay branching ratios $\mathbf{BR_Z}$ are defined in the 
same manner as $\mathbf{BR_H}$.  The flavor tagging efficiency is assumed
 to be largely
independent of whether a jet originated from a Higgs decay or from a $Z^0$
decay.  This assumption may need to be revised when considering Higgs 
masses much larger than the $Z^0$ mass.  

Neglecting the error in the efficiency matrix (measured by counting the
 mixed tag rates where one jet is tagged as flavor $i$ and the other as
flavor $j$) and the error from 
subtracting the $Z^0$ background (presumably measured from the shape of the
recoil mass distribution), the branching ratio error may be estimated as:
\begin{equation}
\Delta \mathbf{BR_H} \approx \mathbf{E}^{-1} \cdot \sqrt{\mathbf{T}} / N_H
\end{equation}
In particular, if the Poisson fluctuations $\Delta T_b$ and $\Delta T_u$ are
ignored then for Higgs to $c\bar c$ decays:
\begin{equation}
\Delta BR_{H\rightarrow c\bar c} \approx E_{cc}^{-1} \cdot \sqrt{T_c} / N_H
\end{equation}
The number of $c$-tagged events is estimated as
\begin{equation}
T_c \approx (E_{cc}\cdot BR_{H\rightarrow c\bar c} +E_{bc}\cdot BR_{H\rightarrow b\bar b})\cdot N_H + E_{cc}\cdot BR_{Z\rightarrow c\bar c} \cdot N_Z 
\end{equation}
where the small contamination from $BR_{Z\rightarrow b\bar b}$ is ignored.
The fractional measurement error is then:
\begin{equation}
\frac{\Delta BR_{H\rightarrow c\bar c}}{BR_{H\rightarrow c\bar c}} \approx 
\frac{\sqrt{E_{cc}\cdot BR_{H\rightarrow c\bar c} +E_{bc}\cdot BR_{H\rightarrow b\bar b} + E_{cc}\cdot BR_{Z\rightarrow c\bar c} \cdot N_Z/N_H}}{E_{cc} \cdot BR_{H\rightarrow c\bar c} \cdot \sqrt{N_H}}
\end{equation}
It is estimated that the Higgs tag has $75\%$ purity and so 
$N_Z/N_H \approx 0.33\%$.  It is also assumed that the true values of the
branching ratios are:  $BR_{H\rightarrow c\bar c}=3\%$, $BR_{H\rightarrow b\bar b}=72\%$, and $BR_{Z\rightarrow c\bar c}=12.4\%$.

In figures~\ref{F:varyhitres}-~\ref{F:varyenergy} below, for
various vertex detector configurations listed in table~\ref{T:trackres},
the efficiency-purity curve of the event flavor tag is plotted. 
These curves are generated by varying the value of the cut on the neural
network output.  The scaled precision $P$, defined as:
\begin{equation}
P = \sqrt{N_H} \cdot \Delta BR_{H\rightarrow c\bar c}/BR_{H\rightarrow c\bar c}
\end{equation}
is also plotted as a function of the event flavor tagging point chosen on the
efficiency-purity curve.  This quantity is a measure of the analyzing power
and can be divided by $\sqrt{N_H}\approx \sqrt{1370}\approx 37$ for
a 500 fb$^{-1}$ data set to get the fractional branching ratio error.

For example, for the standard SDMAR01 configuration with 5 $\mu$m hit
resolution and $0.12\% \ X^0$ thickness and inner radius 1.2 cm, 
figure~\ref{F:varyhitres} indicates that the
best precision occurs at $E_{cc}\approx 0.66$ and $E_{bc} \approx 0.06$ and
the relative branching fraction error can be calculated to be  
$\Delta BR_{H\rightarrow c\bar c}/BR_{H\rightarrow c\bar c} \approx 13.8/37 \approx 0.37$.  
For $N_H\approx 1370$ and $N_Z\approx 452$, the breakdown of the tagged
charm events is $\sim 27$ correct tags from $H^0\rightarrow c\bar c$,
$\sim 59$ mistags
from $H^0\rightarrow b\bar b$, and $\sim 36$ correct background tags from
$Z^0\rightarrow c\bar c$.  The mistags from $Z^0\rightarrow b\bar b$ are
neglected.
Our result is very consistent with that of the Oregon result\cite{Brau:2000dq}
of $\Delta BR_{H\rightarrow c\bar c}/BR_{H\rightarrow c\bar c} \approx 0.39$ for the same model.

\section{The Results}
The single hit resolution $\sigma_{res}$ is varied 
(figure~\ref{F:varyhitres}) between 
$1 \mu$m and $10 \mu$m to cover resolutions ranging from CCD
pixel detectors and of silicon strip detectors.  For example, changing 
from the $5 \mu$m resolution SDMAR01 detector
to a $1 \mu$m detector only decreases the error by a factor of
$13.3/13.8=0.964$.  The changes in precision may be estimated as
$\Delta{P}/\Delta{\sigma_{res}} \approx 0.12 \ \mu$m$^{-1}$.

\begin{figure}[htb]
 \centering
 \begin{minipage}{.47\textwidth}
 \resizebox{\textwidth}{!}{\includegraphics{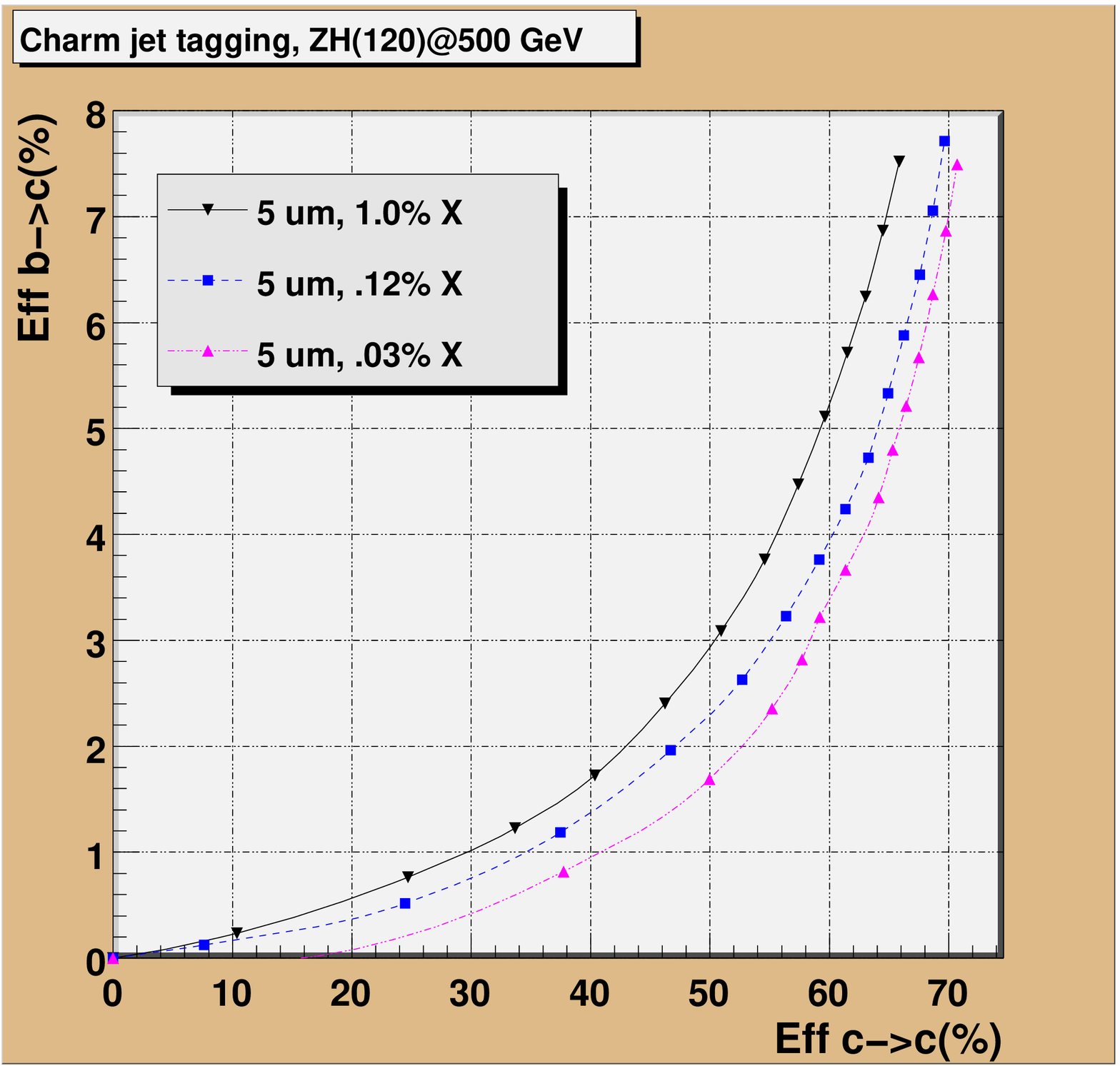}}
 \end{minipage}
 \hfill
 \begin{minipage}{.47\textwidth}
 \resizebox{\textwidth}{!}{\includegraphics{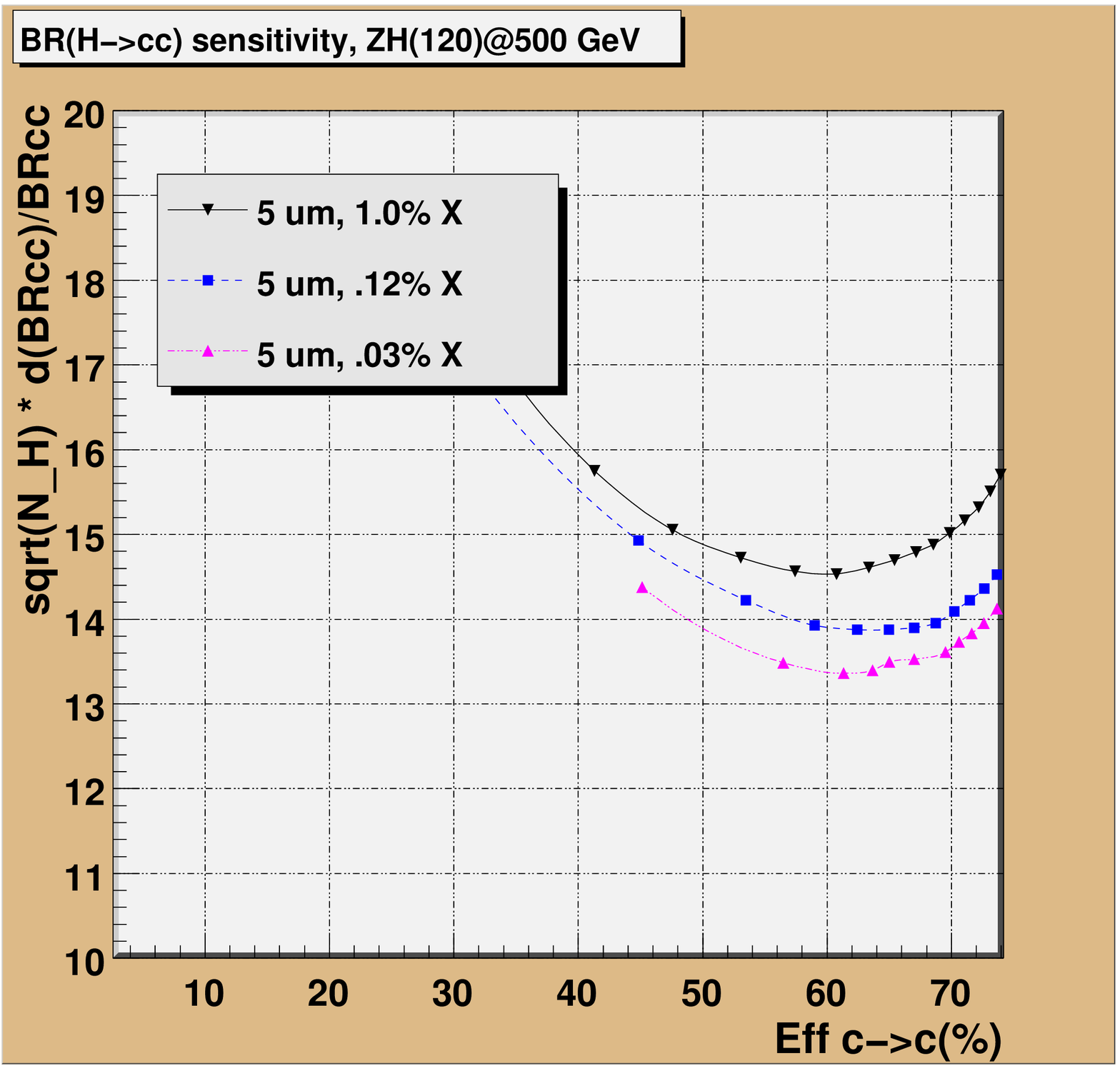}}
 \end{minipage} 
 \caption{\label{F:varythickness} (Left) The event mistag rate $E_{bc}$ versus the event correct tag rate $E_{cc}$  for vertex detectors of different layer thickness but the same hit resolution of 5 $\mu$m and the same inner radius of 1.2 cm.  (Right) The corresponding fractional $BR(H\rightarrow c\bar c)$ precision.}
 \end{figure}

\begin{figure}[htb]
 \centering
 \begin{minipage}{.47\textwidth}
 \resizebox{\textwidth}{!}{\includegraphics{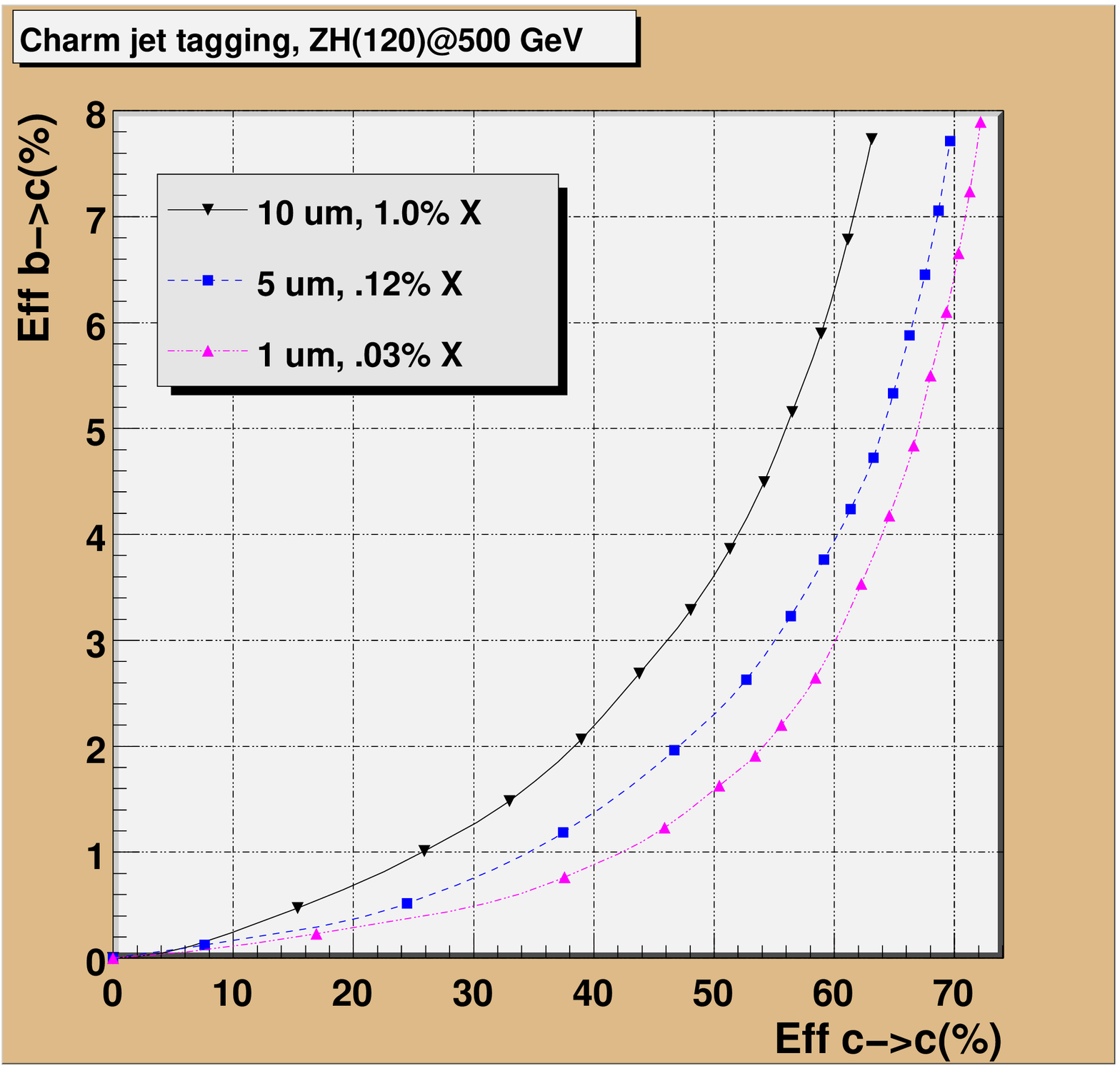}}
 \end{minipage}
 \hfill
 \begin{minipage}{.47\textwidth}
 \resizebox{\textwidth}{!}{\includegraphics{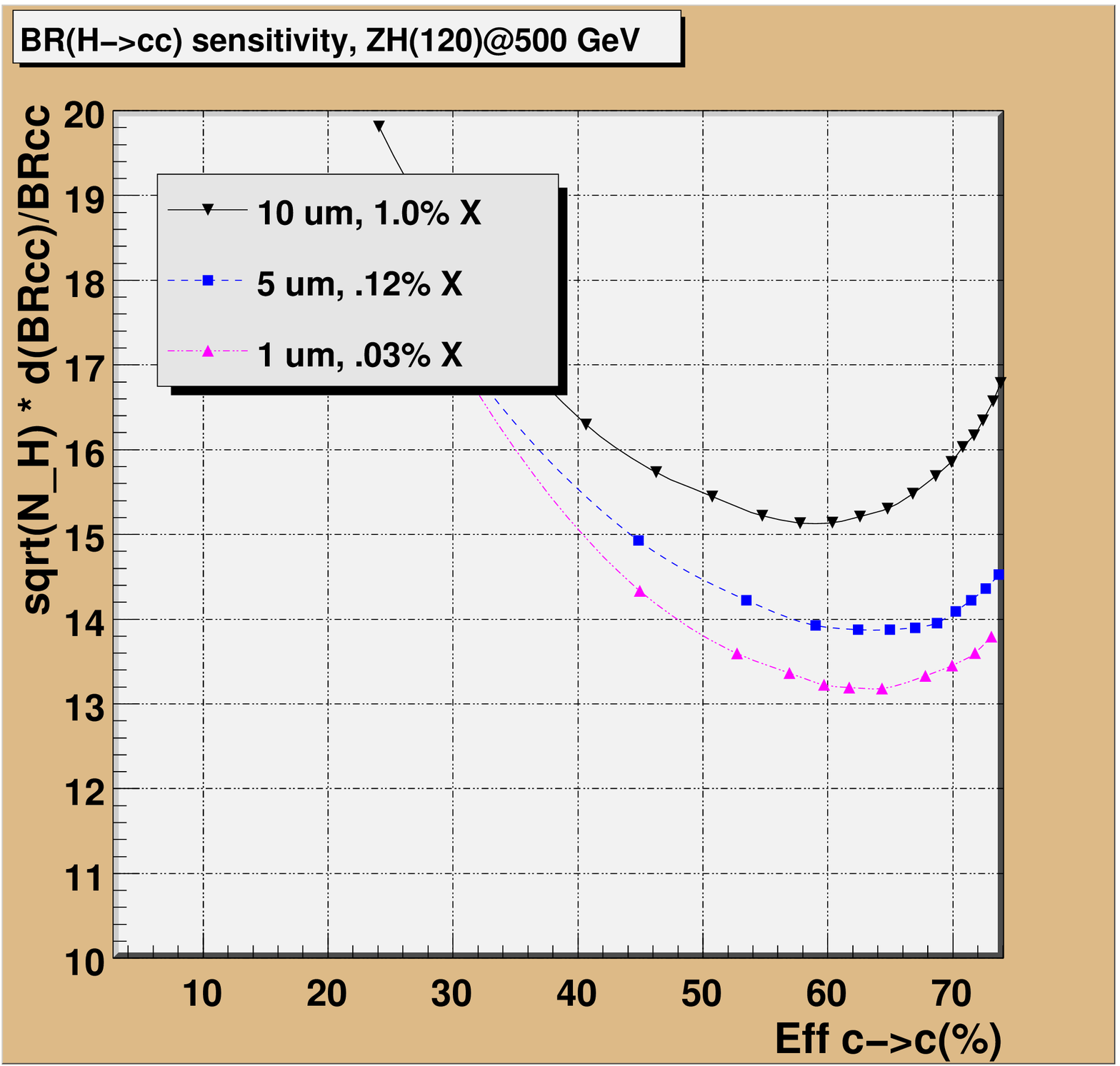}}
 \end{minipage} 
 \caption{\label{F:varyboth} (Left) The event mistag rate $E_{bc}$ versus the event correct tag rate $E_{cc}$  for vertex detectors of different layer thickness and different hit resolution but the same inner radius of 1.2 cm.  (Right) The corresponding fractional $BR(H\rightarrow c\bar c)$ precision.}
 \end{figure}

A large range of variations is chosen for the detector layer thickness $t$,
from $0.03\% X^0$ to $1.00\% X^0$ again to model anything from the thinnest
CCD structures to silicon strips (with readout electronics electronics on
the ends).  These variations, shown in figure~\ref{F:varythickness} resulted
again in changes of precision of magnitude similar to those of the single hit
resolution variations.  These changes may be roughly parameterized as
$\Delta{P}/\Delta{\sqrt{t}} \approx 1.38$ where $t$ is measured in $\% X^0$.
The larger effects of simultaneous variations in both single hit resolution and
layer thickness are shown in figure~\ref{F:varyboth}.  

\begin{figure}[htb]
 \centering
 \begin{minipage}{.47\textwidth}
 \resizebox{\textwidth}{!}{\includegraphics{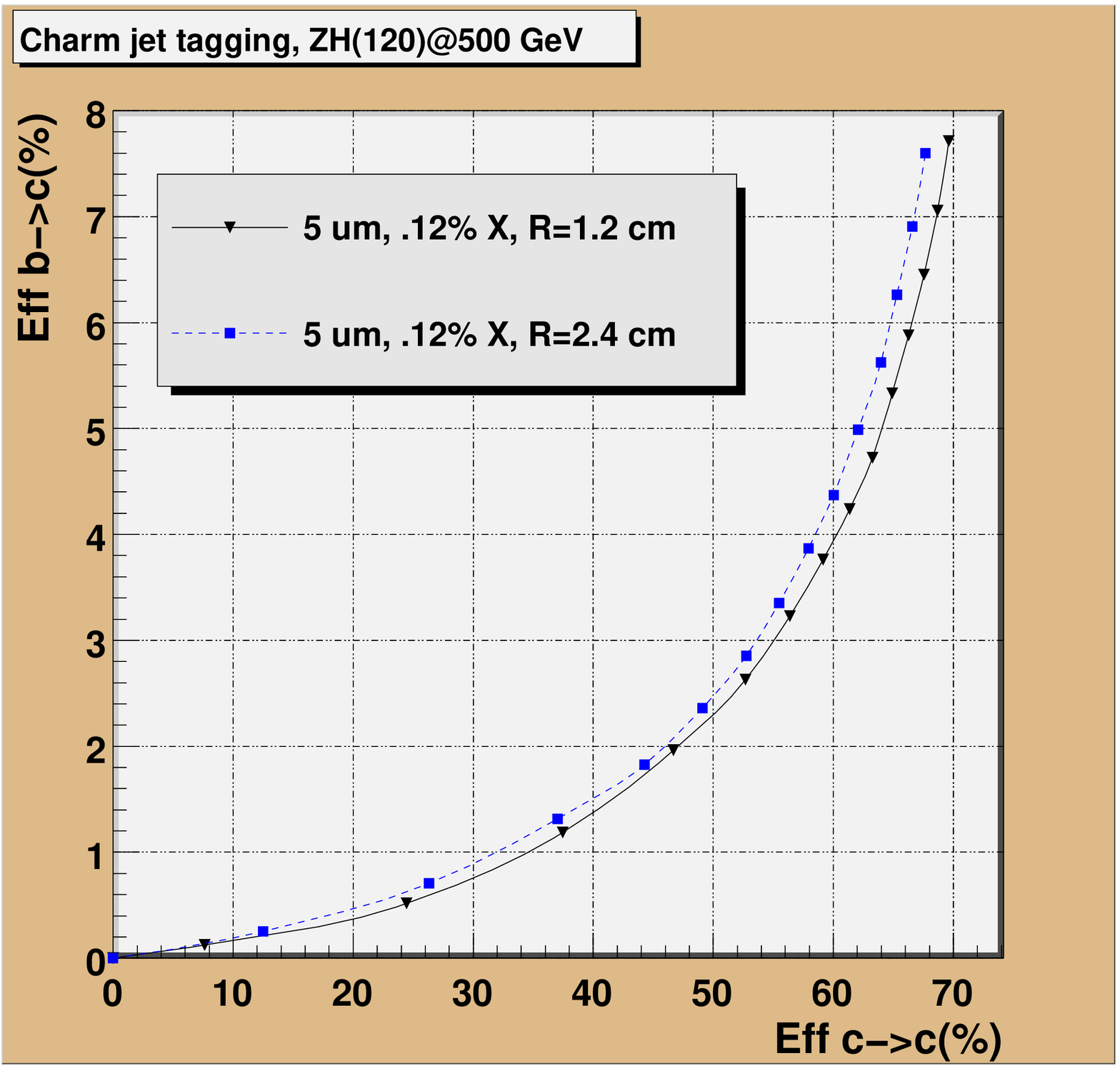}}
 \end{minipage}
 \hfill
 \begin{minipage}{.47\textwidth}
 \resizebox{\textwidth}{!}{\includegraphics{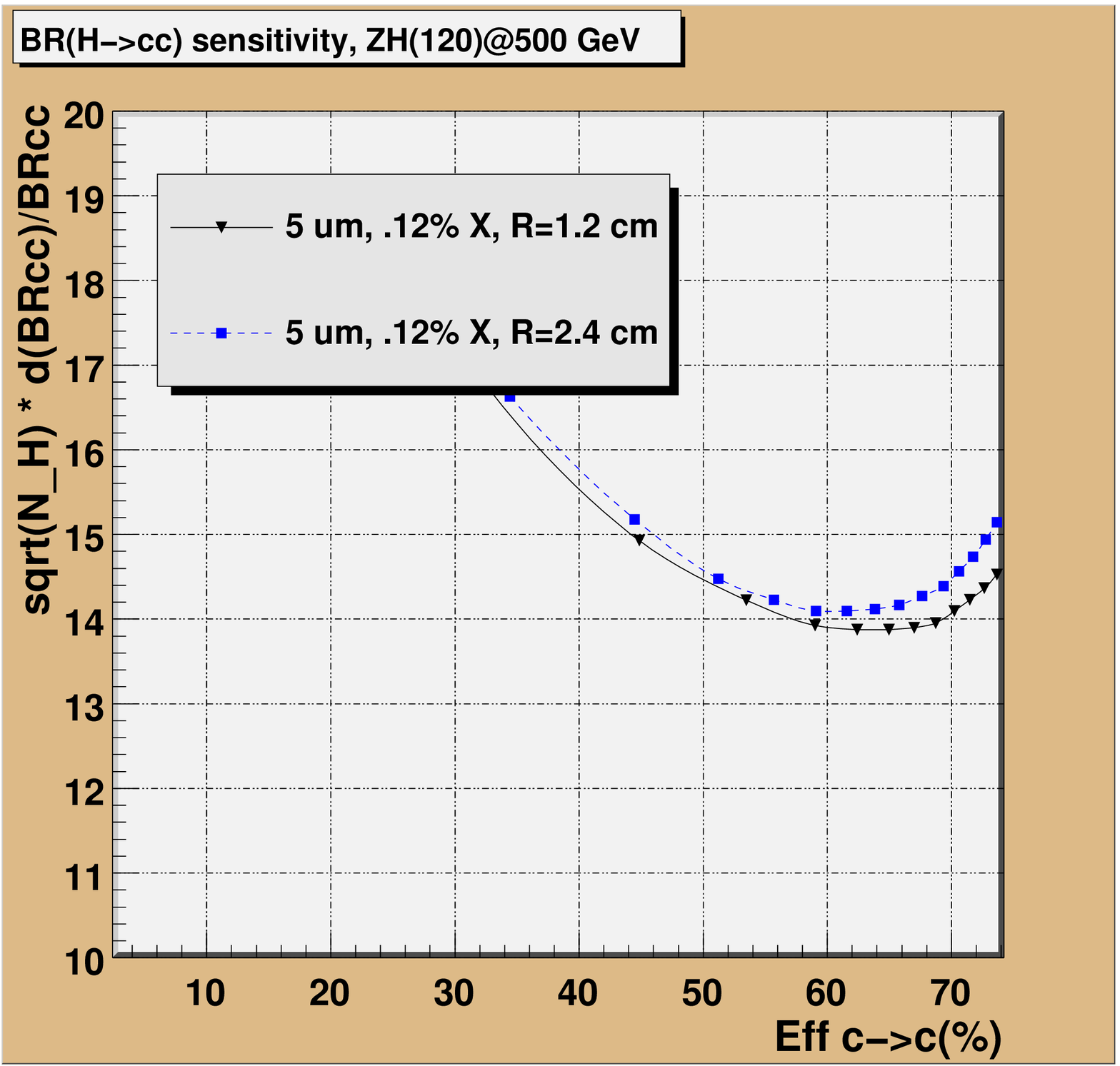}}
 \end{minipage} 
 \caption{\label{F:varyr0} (Left) The event mistag rate $E_{bc}$ versus the event correct tag rate $E_{cc}$  for vertex detectors with different inner radius, using the same hit resolution of 5 $\mu$m and the same layer thickness of $0.12\% \ X^0$.  (Right) The corresponding fractional $BR(H\rightarrow c\bar c)$ precision.}
 \end{figure}

The dependence on the vertex detector inner radius $R_0$ is shown in
figure~\ref{F:varyr0}.  Not much difference can be seen, and the dependence
is approximately $\Delta{P}/\Delta{R_0} \approx 0.17/$cm.  This result
indicates that it may be possible to increase the detector inner radius
to better avoid the $e^+e^-$ pair backgrounds without sacrificing much
precision in the branching ratio measurement.

\begin{figure}[htb]
 \centering
 \begin{minipage}{.47\textwidth}
 \resizebox{\textwidth}{!}{\includegraphics{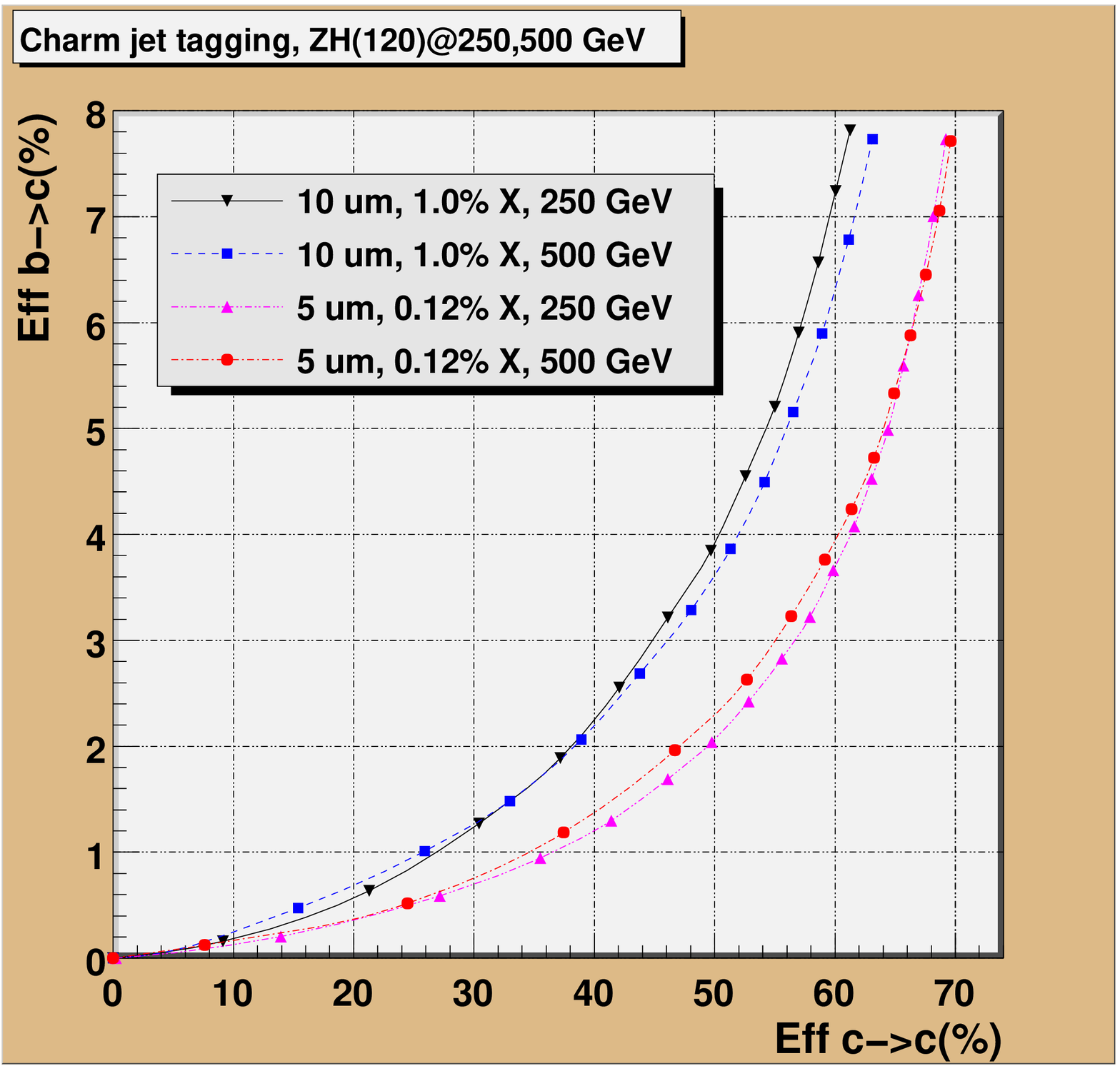}}
 \end{minipage}
 \hfill
 \begin{minipage}{.47\textwidth}
 \resizebox{\textwidth}{!}{\includegraphics{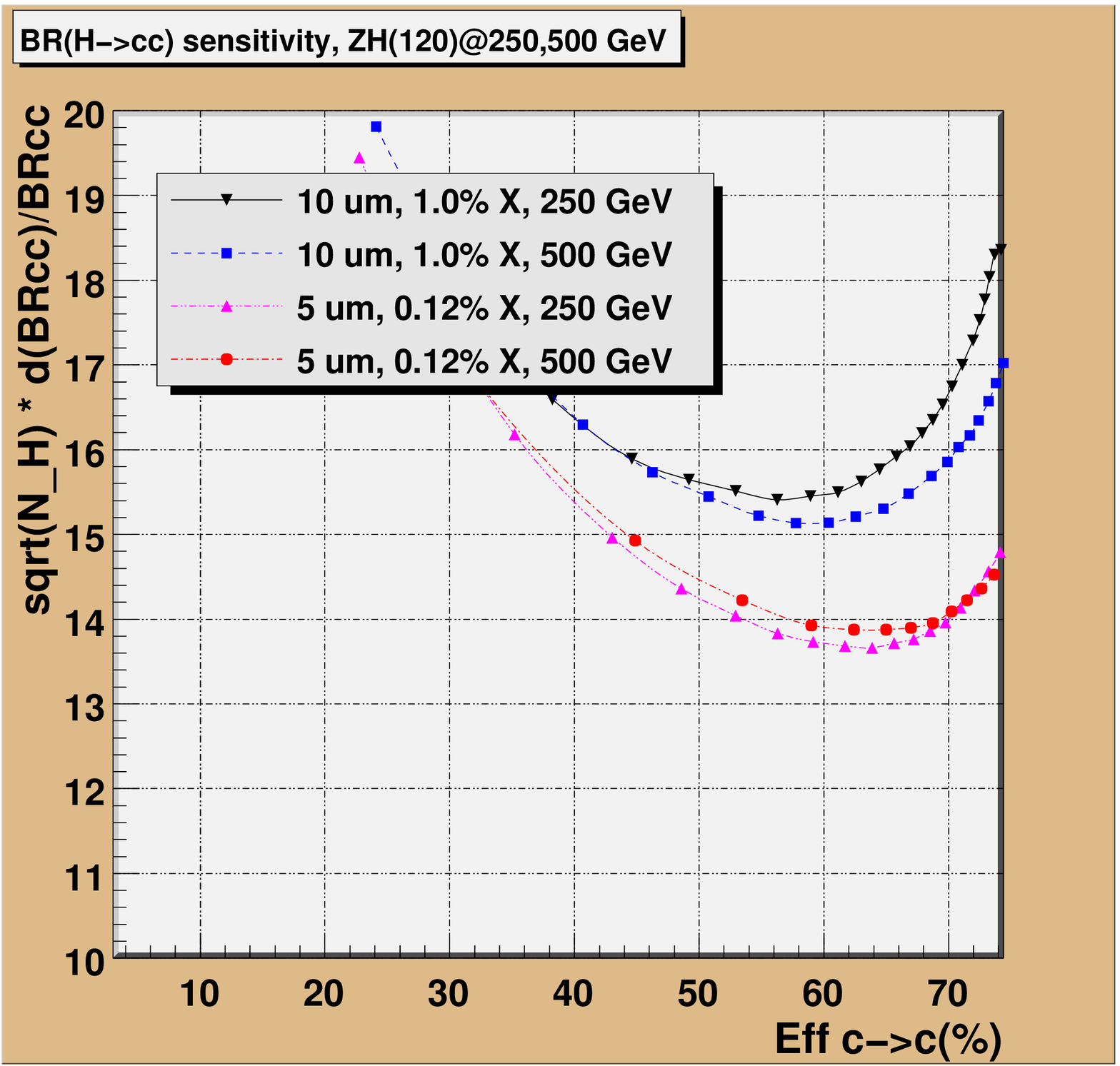}}
 \end{minipage} 
 \caption{\label{F:varyenergy} (Left) The event mistag rate $E_{bc}$ versus the event correct tag rate $E_{cc}$  at two different $\sqrt{s}$ for two detector models.  The inner radius is kept at 1.2 cm.  (Right) The corresponding fractional $BR(H\rightarrow c\bar c)$ precision.}
 \end{figure}

Finally, the dependence of the measurement on the center-of-mass energy
is shown in figure~\ref{F:varyenergy} for $\sqrt{s}= 250, 500$ GeV for
two different detector configurations.  Changing the energy appears to
have only small effects on the flavor tagging contribution to
the measurement precision.  Perhaps more important are the effects of
different Standard Model backgrounds to the Higgs tag at the two different
energies which have not been modelled in this study.

\section{Conclusion}
In this study, we have found that the precision of a
$BR(H^0\rightarrow c\bar c)$ measurement is only modestly dependent on the 
variations of several detector parameters.  This small dependence can
be explained by the fact that the base configuration (SDMAR01) already
provides exceedingly good tracking resolution.  Even the worst 
detector configuration modelled offers resolution much better than that of any 
existing vertex detector.  The modelled tracking resolutions (shown in
table~\ref{T:trackres}) are much smaller than the distance scales given
by the lifetimes of weakly decaying heavy hadrons, for example
c$\tau_D\approx 140 \mu$m.  Flavor tagging analyses are therefore able
to cut deeply into the exponential impact parameter distribution to achieve
a high efficiency for separating heavy flavor decay tracks from 
fragmentation tracks.  Further improvements in detector precision can
only hope to recover the small remaining fraction of the exponential
distribution at low proper time.  

It therefore does not seem necessary from the standpoint of flavor
tagging to attempt to go to extraordinary lengths to achieve further
improvements in resolution.  However, other applications of precision
tracking still need to be studied.  For example, in measurements of
parity violation or CP violation in new physics sectors or even at
the $Z^0$ pole in a giga-$Z^0$ run, it is
necessary to determine not just the particle species, but also its
charge.  Even if a newly discovered particle does not have a weak decay 
scale lifetime, its charge may be measured if it decays to a $b$ or $c$
hadron.  In either case, the loss of even a single secondary decay track
creates an impurity in the charge measurement.  It is therefore
expected that these types of measurements may be more sensitive to
changes in the tracking resolution.

 \begin{figure}[htb]
 \centering
 \begin{minipage}{.47\textwidth}
 \resizebox{\textwidth}{!}{\includegraphics{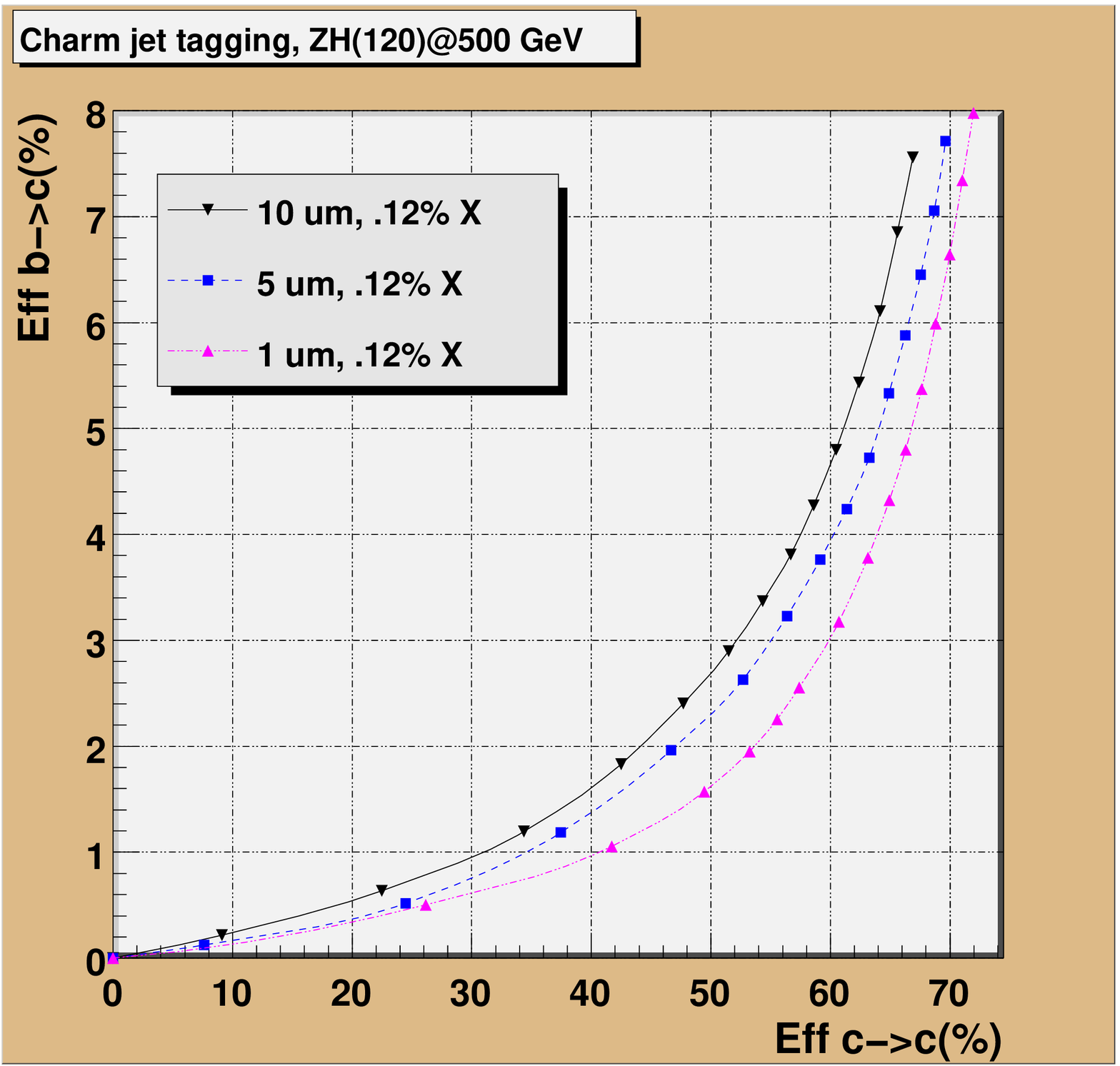}}
 \end{minipage}
 \hfill
 \begin{minipage}{.47\textwidth}
 \resizebox{\textwidth}{!}{\includegraphics{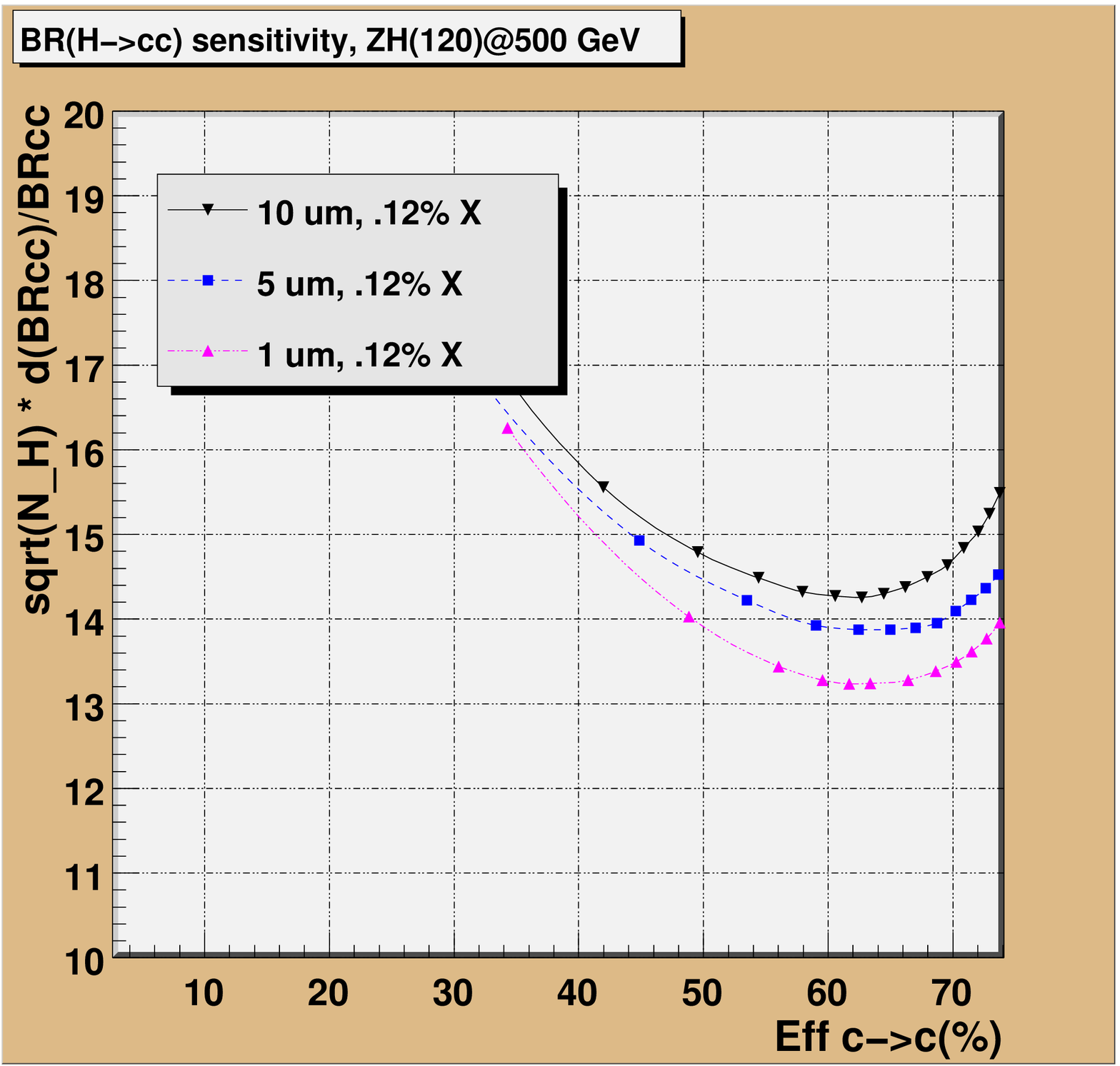}}
 \end{minipage} 
 \caption{\label{F:varyhitres} (Left) The event mistag rate $E_{bc}$ versus the event correct tag rate $E_{cc}$  for vertex detectors of different hit resolution but the same layer thickness of $0.12\% \ X^0$ and the same inner radius of 1.2 cm.  (Right) The corresponding fractional $BR(H\rightarrow c\bar c)$ precision.}
 \end{figure}

\begin{table}
\begin{tabular}{|r|r|r|r|r|}
\hline
hit res. [$\mu$m] & thickness [\% $X_0$] & inner radius [cm] & $\sigma_0$ [$\mu$m] & $\sigma_{MS}$ [$\mu$m$\cdot$GeV/c] \\
\hline\hline
10.0 & 0.12 & 1.2 & 4.7 & 11.7 \\
5.0  & 0.12 & 1.2 & 2.5 & 8.0  \\
1.0  & 0.12 & 1.2 & 0.7 & 4.8  \\
5.0  & 1.00 & 1.2 & 2.5 & 16.0 \\
5.0  & 0.03 & 1.2 & 2.5 & 6.0  \\
10.0 & 1.00 & 1.2 & 4.7 & 20.4 \\
1.0  & 0.03 & 1.2 & 0.7 & 3.1  \\
5.0  & 0.12 & 2.4 & 2.6 & 12.7 \\
(SLD) 4.0  & 0.40 & 2.8 & 9.0 & 33.0 \\
\hline
\end{tabular}
\caption{\label{T:trackres} The different detector configurations based on variations of SDMAR01 and the corresponding track position resolution, $\sigma_0 \oplus \sigma_{MS}/(P \cdot \sin^{3/2}\theta)$.  The last entry represents the SLD model which has an enhanced $0.65\% \ X^0$ beampipe at 2.5 cm radius and has the SDMAR01 tracking chambers replaced with a model of the SLD vertex detector and central drift chamber.  All other models have a $0.04\% \ X^0$ beampipe at 1.0 cm radius.}
\end{table}  

\bibliography{hcc}

\end{document}